\documentclass[twocolumn,a4paper]{article} 

\RequirePackage{amsmath}[1997/03/20]
\RequirePackage{amssymb}
\RequirePackage{graphicx}

\title{The Pierre Auger Observatory: Results on Ultra-High Energy Cosmic Rays}

\author{
Johannes Bluemer\\ 
Karlsruhe Institute of Technology KIT\footnotemark  \\
Postfach 3640, D-76021 Karlsruhe, Germany\\
\and
for the Pierre Auger Collaboration  \\
Observatorio Pierre Auger, Av.  San Martin Norte 304, 5613
Malargue, Argentina
}

\begin{document}

\maketitle

\begin{center}
    {\footnotesize Proceedings of the International Workshop on Advances in
Cosmic Ray Science, Waseda University, Shinjuku, Tokyo, Japan, March
2008; to be published in the Journal of the Physical Society of Japan
(JPSJ) supplement.}
\end{center}

\abstract{The focus of this article is on recent results on ultra-high
energy cosmic rays obtained with the Pierre Auger Observatory.  The
world's largest instrument of this type and its performance are
described.  The observations presented here include the energy
spectrum, the primary particle composition, limits on the fluxes of
photons and neutrinos and a discussion of the anisotropic distribution
of the arrival directions of the most energetic particles.  Finally,
plans for the construction of a Northern Auger Observatory in
Colorado, USA, are discussed.}


\footnotetext{KIT is the cooperation of Universit\"at Karlsruhe (TH)
and Forschungszentrum Karlsruhe GmbH}

\section{Introduction}
The Earth's atmosphere is exposed to a flux of energetic particles
from space.  Their energies extend from the MeV range to at least
$10^{20}$~eV. The non-thermal energy spectrum follows a power law
$dN/dE\propto E^\gamma$; the spectral index $\gamma$ has a value
around $-3$; it varies in certain energy regions indicating
interesting possible changes in the composition and in the
acceleration and propagation processes.

Diffusive shock acceleration is thought to be the basic mechanism that
gives energy to charged cosmic ray particles.  The maximum attainable
energy is proportional to the particles' charge and to the product of
shock velocity, magnetic field and size of the acceleration region
\cite{Hillas1984}.  Supernova remnants are candidates for the
acceleration of galactic cosmic rays up to energies of several
$10^{17}$ eV. Of particular interest is the study of cosmic particles
at the highest energies observed so far.  As cosmic rays of energy
greater than $~10^{19}$\,eV are not confined by typical galactic
magnetic fields, it is natural to assume that those are produced by
extra-galactic sources.  The list of the very few viable candidate
sources for such energies includes active galactic nuclei (AGN), radio
lobes of FR II galaxies, and gamma-ray bursts (GRBs).  However, the
short energy loss lengths indicate that particles of energy above
$~10^{20}$\,eV should come from sources within a $~100$\,Mpc sphere.
Astrophysical sources within our Galaxy are disfavoured.  For a recent
review of astrophysical sources, see \cite{Torres2004}).

\begin{figure}
\begin{center}
 \includegraphics[width=0.9 \columnwidth]{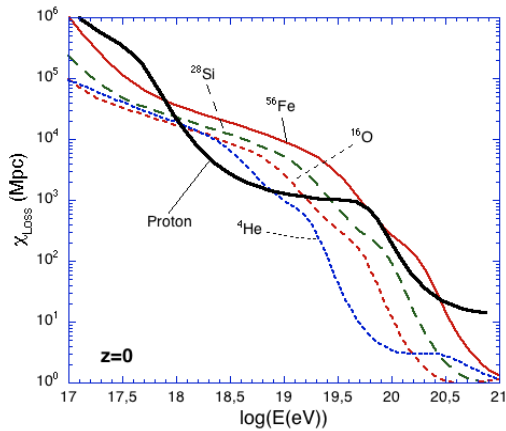}
\end{center}
\caption{ Energy loss lengths \cite{Allard2006} of protons and nuclei
calculated for a redshift $z=0$.  Iron nuclei have larger or similar
attenuation lengths to protons up to $3\times 10^{20}$ eV.}
\label{fig:eloss}
\end{figure}

More than 40 years ago Greisen, Zatsepin and Kuzmin realized that the
interaction of protons with cosmic microwave background photons would
result in significant energy losses.  The energy spectrum would show a
flux suppression above a threshold energy of about $6\times
10^{19}$~eV, the GZK effect \cite{gzk66}.  Similarly, heavy nuclei are
broken up due to photo-disintegration.  A compilation of energy loss
lengths of protons and heavier nuclei is shown in
Figure~\ref{fig:eloss} \cite{Allard2006}.  Since the energy loss
processes for protons and nuclei are different the evolution of
composition during propagation has to be taken into account.

Alternative, non-acceleration scenarios for ultra-high energy cosmic
rays have been proposed, such as the decays of topological defects or
super-heavy dark matter.  All of these models predict high fluxes of
gamma rays and neutrinos.  Finally there are propagation models in
which the GZK energy loss processes are evaded or shifted to higher
energies.  Examples are violation of Lorentz invariance, the $Z$-burst
model or postulation of new particles with properties similar to
protons.  Reviews of such alternative scenarios can be found in 
\cite{Bhattacharjee2000, Kachelriess2004}.  They are now disfavoured
except for the highest energies due to already stringent limits on the
flux of photons.

Measurements of the arrival direction distribution, primary mass
composition and energy spectra are the keys to solving the puzzle of
ultra-high energy cosmic particles.  High statistics is particularly
important at the highest energies, where magnetic deflections are
minimal and even charged particles are expected to point back to their
source regions.


\section{The Pierre Auger Observatory}

The Pierre Auger Observatory for the highest-energy cosmic rays has
been developed and built by more than 350 scientists in 17 countries.
The southern site in Mendoza, Argentina, will be completed during the
year 2008.  The Observatory comprises 1600 water-Cherenkov detectors
deployed over 3000 km$^{2}$ on a triangular grid with 1500 m spacing.
This surface detector array (SD) is overlooked by 24 electronic
telescopes, which are arranged in four stations around the SD area.
The telescopes record images of the faint, ultra-violet fluorescence
light excited by the air showers.  The fluorescence detector (FD)
operates on clear dark nights and achieves a duty cycle of 13 percent.
The properties and performance of the prototype instruments have been
published in 2004 \cite{eanim}.  A recent description of the combined
hybrid operation of the SD and FD systems of the Auger Observatory can
be found in the proceedings of the ICRC 2007 \cite{icrcdawson}.

\begin{figure}
\begin{center}
 \includegraphics[width=0.9 \columnwidth]{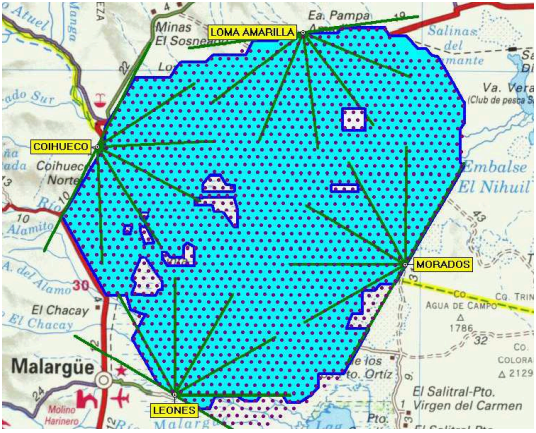}
\end{center}
    \caption{The southern Pierre Auger Observatory is located in the
    El Nihuel region in the Province of Mendoza, Argentina; the
    central campus is in the city of Malargue.  Dots indicate the 1600
    surface detector positions.  The four named fluorescence detector
    stations with six individual telescopes each are all operational,
    as well as the surface detectors within the shaded area.}
    \label{fig:auger-array}
\end{figure}

The layout and deployment status as of June 2008 is shown in Figure
\ref{fig:auger-array}.  The data set underlying the results reported
here has been collected basically from January 2004 to August 2007.
During this time the SD system grew from 154 to 1200 tanks and the FD
system from 6 to 24 telescopes.  The exposure over this period is
three times greater than that of AGASA, similar to the monocular HiRes
exposure, and corresponds to about 80 percent of a full Auger-year.
Above 1 EeV the Auger Observatory has recorded more events than all
previous efforts together.  Due restrictions to periods of good data
quality have been applied.  We propose to use the unit Linsley as a
measure of the total exposure of cosmic ray observatories, $1 Linsley
= 1 km^{2} sr yr$.  At the time of writing (June 2008) the total Auger
exposure is $10,000 L$.  The expected future annual increment is 7000
L.

From the beginning the Auger Observatory had been conceived as a
hybrid system, in which the SD and FD components are completely
integrated.  The merits of the SD include stable operation with 100
percent duty cycle and a relatively straight-forward determination of
the effective area times solid angle (the aperture), whereas the
energy measurement has to rely heavily on Monte Carlo simulations.
Complementary, the FD offers optical shower detection in a
calorimetric way and can be calibrated with very little dependance on
shower models; however, the aperture grows with shower energy and its
determination for all operating conditions is nontrivial.

A public internet display of 1\% of the events recorded by the Auger
Observatory is available online at http://augersw1.physics.utah.edu/ED/.

Each SD station is a 3.6 m diameter polyethylene tank containing a
sealed liner with a reflective inner surface.  The liner contains
12,000 l of pure water.  Cherenkov light produced by the passage of
particles through the water is collected by three nine-inch-diameter
photomultiplier tubes, which are symmetrically distributed at a
distance of 1.20 m from the center of the tank and look downwards
through windows of clear polyethylene into the water.  The surface
detector station is autonomously operating with a 10 Watt solar power
system.  The electronics package includes a processor, GPS receiver,
radio transceiver and power controller.

A vertically arriving cosmic ray shower of $10^{19}$ eV typically
triggers 8 detectors.  SD signals are measured in units of a
relativistic muon passing vertically through the centre of a
water-Cherenkov detector (vertical equivalent muon, VEM).  This
calibration of the SD array is performed continuously during
data-taking with 2\% accuracy.  The number of secondary particles at
observation level increases linearly with primary energy with some
weak dependencies on the primary mass.  The measured signal as a
function of distance from the shower axis is fitted for individual
events to obtain the VEM particle density at 1000m, S(1000), which is
used as an estimator of the size of each event.  For energies above
$10^{19}$ eV, the relative uncertainty of S(1000) is about 10\%,
including contributions from counting statistics, from the imperfectly
known lateral particle density distribution and from shower-to-shower
fluctuations.

At least five active stations must surround the detector with the
highest signal, and the reconstructed shower core must lie inside an
active triangle of detectors.  For most analyses we use showers up to
zenith angles $\theta < 60$ degrees.  These quality cuts result in
full acceptance above 3 EeV, and nearly constant exposure as function
of $sin^{2}\theta$.  The effective area of the entire array at any
time has been calculated from the number of active hexagons, which can
be deduced from the low- level triggers sent by each detector every
second.  The integrated exposure is 7000 $km^{2}$ sr yr and is known
to 3\%.  Periods corresponding to less than 10\% of the integrated
exposure had data acquisition problems and were removed.

The angular resolution of the surface detector was determined
experimentally, checked using the pairs data set and found to be
better than 2$^{\circ}$ for 3-fold events ($E < 4$ EeV), better than
1.2$^{\circ}$ for 4-fold and 5-fold events ($3 < E < 10$ EeV) and better
than 0.9$^{\circ}$ for higher multiplicity events, which have more than
10 EeV.  

The FD measures the longitudinal development of cosmic ray showers in
the atmosphere.  Each individual telescope of the FD images a portion
of the sky of 30 degrees in azimuth and elevation.  Light is collected
by a segmented spherical mirror of 3.6 x 3.6 $m^{2}$ through a
UV-transparent filter window and a ring corrector lens to reduce the
abberations inherent in the Schmidt optics.  The camera consists of
440 hexagonal photomultipliers, each with a field of view of 1.5 in
diameter.  The signals are continuously digitized at 10 Ms/s,
temporarily buffered and searched for shower track patterns in real
time.  The timing of FD and SD systems are synchronized to about 120
ns.  The data are merged offline in the event building process.

The fluorescence light produced by the shower is detected as a line of
pixels in the FD camera.  The plane defined by the orientation of
pixels, together with the timing information, is used to determine the
shower direction.  The pixel signal must be corrected for attenuation
of the fluorescence light due to Rayleigh and aerosol scattering along
its path toward the telescope, and for a contribution of direct and
scattered Cherenkov light.  The fluorescence light is then
proportional to the electromagnetic energy deposited by the shower
along its path in the atmosphere.  The reconstructed energy profile is
fitted with a Gaisser-Hillas function \cite{gaisser-hillas}, which
provides a measurement of the maximum depth of the shower, and of the
shower energy.  A final correction for missing energy due to high-energy
muons and neutrinos amounts to about $(10\pm4)\%$.  This procedure
provides a nearly calorimetric, model-independent energy measurement
with a resolution of 8\%.

The absolute fluorescence yield of the 337 nm band in air is 5.05
photons/MeV of energy deposit at 293 K and 1013 hPa, derived from
\cite{nagano2004}.  We use the measurements of \cite{airfly2007} for
the wavelength and pressure dependence of the fluorescence spectrum.

The absolute calibration of the FD telescopes is known to $10\%$.  The
relative response of all FD channels is determined twice every night
illuminating the cameras from pulsed LEDs and/or Xe flashers.  The
absolute calibration is performed less frequently.  In this case the
whole aperture is illuminated by a flat-field source with known
spectral and directional characteristics and known intensity,
calibrated at the National Institute of Standards and Technology.
Atmospheric monitoring is an integrated part of the FD operation.  The
system consists of four single wave-length LIDAR stations, one Raman
LIDAR, several weather stations, balloon born meteorological probes,
remote-controlled laser and dedicated aerosol monitors.

The systematic uncertainties in setting the FD energy scale sum to
22\%.  The largest uncertainties are due to the absolute fluorescence
yield (14\%), the absolute calibration of the FD (10\%) and the
reconstruction method (10\%), while the influence of pressure,
humidity and temperature, the wavelength dependent response of the FD,
the aerosol phase function, missing energy and others are smaller.

Instrumental enhancements are currently being installed close to the
Coihueco FD station.  These include underground muon detectors,
additional water Cherenkov detectors, high-elevation fluorescence
telescopes for a larger field-of-view and radio antenna to detect the
geo-synchrotron emission of air showers.


\section{The Energy Spectrum}

Combining the SD and FD information, the energy can be estimated for
each event by a method that relies almost entirely on data and depends
very little on hadronic interaction models or assumptions about the
nature of the primary particle.  We use the so-called \emph{constant
intensity cut} method \cite{Hersil1961}, by which the zenith angle
dependence of S(1000) is derived from the assumption that the true
cosmic ray intensity at a given energy should be the same for all
directions.  Using a suitably chosen reference intensity we normalize
S(1000) for each event to the signal it would have had at a zenith
angle of 38$^{\circ}$, which is the median zenith angle of the events of
interest.  That quantity called $S_{38}$ is used to correlate with the
calorimetric energy measurement by the FD for currently 661 selected
high-quality hybrid events.  To avoid efficiency and selection biases,
low energy events are discarded from the calibration sample.  The
fitted power law $E_{FD}=1.49\times 10^{17} eV\times S_{38}^{1.08}$
indicates that $S_{38}$ grows approximately linear with energy.  The
energy resolution estimated from the root mean square deviation of the
distribution is 19\%, which is in good agreement with the quadratic
sum of the SD and FD energy uncertainties.

\begin{figure}
\begin{center}
 \includegraphics[width=0.9 \columnwidth]{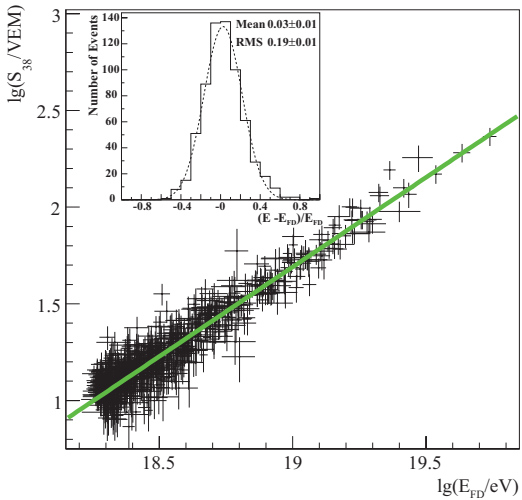}
\end{center}
\caption{Correlation between surface detector signal and FD energy for
661 hybrid events used in the fit.  The full line is the best fit to
the data; the fractional differences between the two energy estimators
are shown in the inset.}
\label{sd-fd-correlation}
\end{figure}

\begin{figure}
\begin{center}
\includegraphics[width=0.9 \columnwidth]{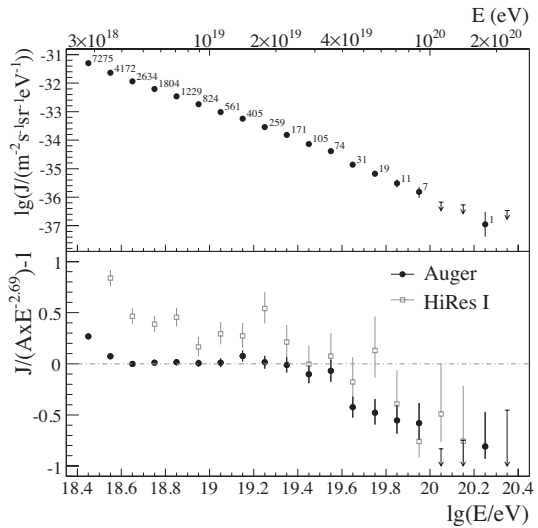}
\end{center}
\caption{Upper panel: Differential flux as a function of energy.
Vertical error bars represent the statistical uncertainty only.  The
number of events in each bin is also given.  Lower panel: Fractional
differences between Auger and HiRes I data relative to a
spectrum with an index of 2.69.}
\label{fig:spectrum}
\end{figure}

The energy spectrum based on 20,000 SD events with the energy scale
set by the FD as described above is displayed in Figure
\ref{fig:spectrum}.  The total systematic energy scale uncertainty is
22\% as described in the previous section.  The residuals relative to
a spectrum with a spectral index of 2.69 is also shown together with
data from the HiRes experiment \cite{abbasi2008}.  No data from the
AGASA experiment are shown as they are currently under revision
\cite{teshima2007}.  The Auger data show the flux suppression above
$4\times10^{19}$ eV with 6 standard deviations.

Statistics and energy range can be extended by including inclined
showers with zenith angles larger than 60$^{\circ}$ and hybrid events,
which have at least one SD detector together with a FD track recorded.
The inclined events add another 1,500 L to the data set at all
energies; special reconstruction efforts are currently being devised
for events with zenith angles larger than 80$^{\circ}$, where the particle
distributions are significantly distorted due to absorption in air
and the Earth's magnetic field. 

\section{The Primary Mass Composition}

\begin{figure}
\begin{center}
    \includegraphics[width=0.9 \columnwidth]{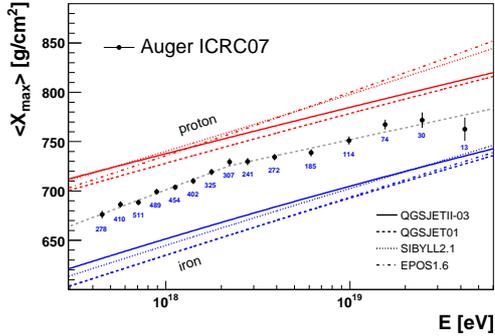} 
\end{center}
\caption{Atmospheric depth $X_{max}$ of the shower maximum as a
function of energy.  The data points are accompanied by their
respective number of events in each bin; the expectations for primary
protons and iron nuclei are shown for three different models
\cite{xmax2007}.}
\label{fig:xmax}
\end{figure}

Measuring the composition of cosmic rays is crucial in order to obtain
a full understanding of their acceleration processes, propagation and
relation with galactic particles.  The atmospheric depth $X_{max}$
denotes the longitudinal position of the shower maximum, which is
directly accessible with the FD. It grows logarithmically with the
energy of the primary particle.  The behaviour of $X_{max}$ for
different primary particles like photons, protons and heavier nuclei
can be conceptually understood in the framework of Heitler models
\cite{matthews2005}, which are in good agreement with detailed Monte
Carlo simulations.  Data and models for Auger hybrid events are shown
in Figure \ref{fig:xmax}.  The appearent change from lighter to
heavier particles above $2\times10^{18}$ eV is an interesting feature,
which is currently under investigation.

\section{The Photon and Neutrino limits}

\begin{figure}
\begin{center}
\includegraphics[width=0.9 \columnwidth]{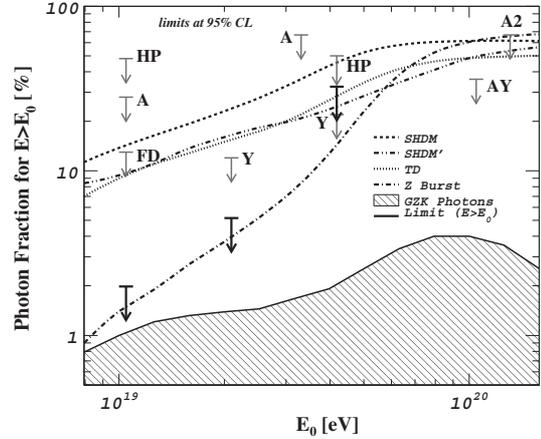} 
\end{center}
\caption{The upper limits on the fraction of photons in the integral
cosmic-ray flux derived from SD events (black arrows) along with
previous experimental limits (HP: Haverah Park; A1, A2: AGASA; AY:
AGASA-Yakutsk; Y: Yakutsk; FD: Auger hybrid limit).  Also shown are
predictions from top-down models and for the GZK photon fraction.  See
\cite{abraham2008-sdphotons} and references therein.}
\label{fig:photon-fraction}
\end{figure}

Photon induced air showers would penetrate 200 - 300 $g/cm^{2}$ deeper
into the atmosphere than protons.  The observables sensitive to the
longitudinal shower development are $X_{max}$ measured with the FD,
and the signal risetime and the curvature of the shower front measured
in SD-only events.  The first limit on the photon contents of cosmic
rays at the highest energies had been derived from the fluorescence
detector \cite{abraham2007-fdphotons}; a more stringent result has now
been derived using the large statistics of surface detector events.
The maximum fraction of photons is 2.0\% above $10^{19}$ eV and 62\%
above $8\times10^{19}$ eV, which corresponds to limits on the flux of
photons of $6.9\times10^{-3}$ and $1.7\times10^{-3} km^{-2} sr^{-1}
yr^{-1}$, respectively, with a confidence level of 95\%.  This new result
\cite{abraham2008-sdphotons} disfavors many exotic models of
sources of cosmic rays as shown in Figure \ref{fig:photon-fraction} .

\begin{figure}
\begin{center}
    \includegraphics[width=0.9 \columnwidth]{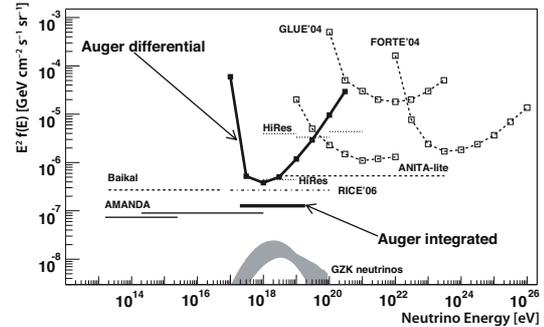} 
\end{center}
\caption{Limits at 90\% C.L. for a diffuse flux of astrophysical
neutrinos.  Limits from other experiments are converted to a single
flavour assuming a 1 : 1 : 1 ratio of the 3 neutrino flavours and
scaled to 90\% C.L. where needed.  The shaded curve shows a typical
range of expected fluxes of GZK neutrinos - predictions vary by an
order of magnitude.  See \cite{abraham2008-nutau} and references
therein.}
\label{fig:neutrino-limits}
\end{figure}

A large air shower array like Auger represents a significant target
mass to high-energy astrophysical neutrinos.  Neutrino induced showers
can be identified if they occur deep in the atmosphere under large
zenith angles, or by their special topology in the case of
Earth-skimming tau neutrinos.  Identification criteria have been
developed to find EAS that are generated by tau neutrinos emerging from
the Earth.  No candidates have been found in the data collected
between 1 January 2004 and 31 August 2007.  We derive an upper limit
on the diffuse tau neutrino flux as $E^{2}_{\nu} dN_{\nu_{\tau}} /
dE_{\nu} < 1.3 \times 10^{-7}$ GeV cm$^{-2}$ s$^{-1}$ sr$^{-1}$ in the
energy range $2 \times 10^{17}$ eV $< E_\nu < 2 \times 10^{19}$ eV.
In Figure \ref{fig:neutrino-limits} we show our result
\cite{abraham2008-nutau}, which is at present the most sensitive bound
on neutrinos in the EeV energy range.  In the future, our sensitivity
will improve by more than an order of magnitude.

\section{The Arrival Directions}

\begin{figure}[h]
\begin{center}
    \includegraphics[width=7.9cm]{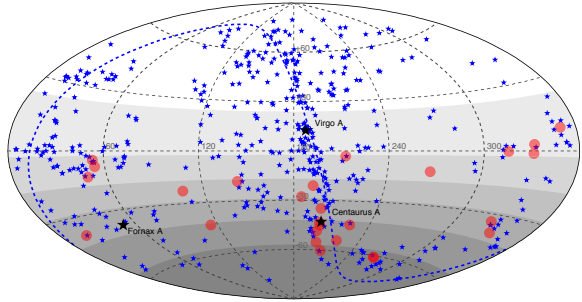} 
\end{center}
\caption{Map of the celestial sphere in equatorial coordinates.
Colored circles with radius 3.1 degrees are centered at the arrival
directions of the 27 cosmic rays with highest energy detected by the
Pierre Auger Observatory.  The positions of the 472 AGN with redshift
z less than 0.018 (corresponding to distances up to 75 Mpc) from the
12th edition of the catalog of quasars and active nuclei by
Veron-Cetty and Veron are indicated by the blue stars.  Grey shading
indicates larger relative exposure, which is maximum at the South
celestial pole.  Centaurus A, Fornax A, and Virgo A, three of the
closest candidate sources, are marked with black stars.  The
super-galactic plane is shown as a dashed blue line.}
\label{auger2007-agn-equatorial}
\end{figure}

The search for anisotropies in the arrival directions of cosmic rays
or even for distinct sources has been a long-standing goal.  The Auger
Collaboration has recently published the observation of a correlation
between the arrival directions of the most energetic cosmic rays and
Active Galactic Nuclei (AGN) listed in the Veron-Cetty and Veron
catalogue \cite{abraham2007-AGNscience, abraham2008-AGNlong}.  An
exploratory scan of an early data set had been performed in the
parameter space given by the angular deviation between arrival
directions and AGN positions, AGN distance (given by their redshift)
and cosmic ray energies.  Parameters were then fixed \textit{a priori}
and applied to an independent second data sample in order to avoid the
complications and statistical penalties inherently present in
\textit{a posteriori} searches.  The correlation has its maximum
significance for energies greater than 57 EeV, AGN closer than about
71 Mpc distance and angular deviations within 3.1$^{\circ}$.  In this
conference contribution a different representation in equatorial
coordinates instead of galactic cooordinates is shown in Figure
\ref{auger2007-agn-equatorial}.

This result has to be considered together with the flux suppression
described previously; it is consistent with the hypothesis that the
rapid decrease of flux measured by the Pierre Auger Observatory above
60 EeV is due to the GZK effect and that most of the cosmic rays
reaching Earth in that energy range are protons from nearby
extragalactic sources, either AGN or other objects with a similar
spatial distribution.  The maximum acceleration power of cosmic ray
sources cannot be determined yet, but this alone cannot explain the
sudden onset of the position correlation.

A rich spectrum of interesting science questions opens up when future
observations of all sources on the sky with high statistics can be
combined with detailed investigations of the primary parcticles'
nature and their interactions at center-of-mass energies up to 350
TeV.

The Auger Collaboration has also searched for signals from the
galactic center \cite{abraham2007-galactic-center}, and for clustering
on different angular scales at the highest energies and for
correlations with BL Lac objects.  These studies \cite{Armengaud2007,
Harari2007, Mollerach2007} have not confirmed previous claims.


\section{The Northern Pierre Auger Observatory}

The scientific challenge is to follow-up the recent results by the
identification of sources over the whole sky and by the detailed study
of the physical processes at work in such extreme conditions.  These
involve strong gravitational fields, high temperatures and densities,
large magnetic and background radiation fields, and very high
energies interactions.  Multiple events from the most intense sources
will allow a detailed study of the source characteristics and
acceleration processes, while particle collisions at center of mass
energies in excess of 300 TeV may reveal new insights to particle
physics.  Such measurements at Auger South have yielded unexpected
results when compared with current interaction models.  High
statistics at the highest energies is essential to address these
scientific objectives effectively.  The Northern Auger Observatory
(Auger North) will focus on achieving higher statistics at energies
above $6 \times 10^{19}$ eV, where the Greisen-Zatsepin-Kuzmin effect
makes it possible to study nearby sources without the isotropic
background from the rest of the far universe.

Auger North will be built with a combination of the same basic
elements as used in Auger South: surface particle detector stations,
fluorescence telescopes, and an associated communications and
calibration infrastructure.  The site, chosen in 2005, is located in
the South-East corner of the State of Colorado (USA), centered at
about 38$^{\circ}$~N~Lat, 102$^{\circ}$~30'~W~Long.  The average
altitude is about 1300 m above sea level.  The landscape is almost
flat, gently rolling and open; it offers an exceptionally large area
of more than 8,000 square miles (20,000 square kilometers), possibly
spanning into the State of Kansas.  An important feature is the system
of county roads, which are spaced on a rectangular one-mile grid
covering a large fraction of the anticipated deployment area.

The conceptual design includes a base grid of 4,000 SD stations to be
positioned on every second corner of the square-mile grid, alternating
in adjacent lines.  The total area is 8,000 miles$^2$ or 20,000 km$^2$
with a detector spacing of approximately $\sqrt{2} \times 1$ mile (2.3
km).  Simulations suggest that this configuration reaches 50$\%$
efficiency at $10^{19}$ eV and 100$\%$ efficiency at $10^{19.5}$ eV.
An embedded denser grid of 400 additional SD stations would be placed
on the empty positions in 10 percent of the base grid in order to
achieve full efficiency already at $10^{19}$ eV for a subset of the
events.  The landscape requires a peer-to-peer data communication
rather than a direct transmission from SD stations to centralized
receivers.  Almost full coverage of the SD system can be achieved with
40 to 50 fluorescence telescopes anticipating a maximum viewing
distance of 40 kilometers.

\begin{figure}
\begin{center}
    \includegraphics[width=0.9 \columnwidth]{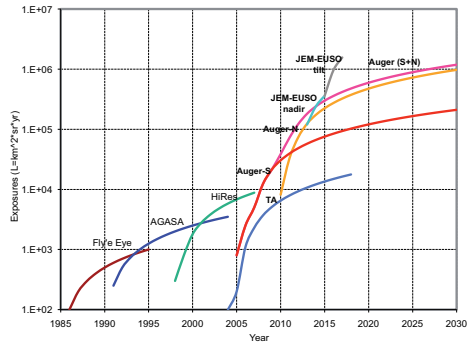} 
\end{center}
\caption{Exposures of cosmic ray observatories as a function of time.
The JEM-EUSO project is a space-born instrument on the ISS, whereas
the other installations are ground-based.
\cite{Olinto2008-exposures}.}
\label{fig:exposures}
\end{figure}

Research and development works in order to improve performance and
reduce cost have started some time ago and activities on the Northern
site in South-East Colorado are ramping up.  Technically, the
construction of Auger North could begin in 2010.  Figure
\ref{fig:exposures} shows the exposures of current and future cosmic
ray observatories as a function of time.

\section{Conclusion and outlook}

The Pierre Auger Observatory has been collecting data of unprecedented
quality since January 2004.  First results include a measurement of
the energy spectrum, which exhibits a flux suppression above 40 EeV as
predicted by Greisen, Zatsepin and Kuzmin.  The arrival directions of
the most energetic cosmic particles are anisotropic and show a
correlation with the positions of nearby extragalactic objects.  Other
previous claims of anisotropies or distinct sources have not been
confirmed.  Preliminary indications of the cosmic ray mass composition
give rise to interesting interpretations.  Based on the wealth of 
current data, a conceptual design for the Northern Auger Observatory 
has been presented.


\section{Acknowledgement}

I would like to thank the organizers of the \textit{International
Workshop on Advances in Cosmic Ray Science} held in March 2008 at
Waseda University, Shinjuku, Tokyo, Japan; it was a pleasure taking
part in the interesting presentations and discussions.  I enjoyed the
help of my Auger colleagues in the preparation of this report.  I
acknowledge the support from Forschungszentrum Karlsruhe, member of
the Helmholtz Association of German Research Centers, and from the
University of Karlsruhe, now joining forces in the \textit{Karlsruhe
Institute of Technology} KIT.


\end{document}